\documentclass[]{aipproc}
\layoutstyle{8x11double}

\usepackage{epsfig}
\usepackage{amsmath}

\begin{document}
\title[Kinematics of GRBs and Afterglows]{Kinematics of Gamma-Ray Bursts and their Relationship to Afterglows}

\author{Jay D. Salmonson}{
	address={Lawrence Livermore National Laboratory, P.O. Box 808,
	Livermore, CA 94551}, email={salmonson@llnl.gov},
	homepage={http://members.home.net/jdsalmonson} }


\copyrightyear  {2001}

\begin{abstract}
A strong correlation is reported between gamma-ray burst (GRB) pulse
lags and afterglow jet-break times for the set of bursts (seven) with
known redshifts, luminosities, pulse lags, and jet-break times.  This
may be a valuable clue toward understanding the connection between the
burst and afterglow phases of these events.  The relation is roughly
linear (i.e.\ doubling the pulse lag in turn doubles the jet break
time) and thus implies a simple relationship between these quantities.
We suggest that this correlation is due to variation among bursts of
emitter Doppler factor.  Specifically, an increased speed or decreased
angle of velocity, with respect to the observed line-of-site, of burst
ejecta will result in shorter perceived pulse lags in GRBs as well as
quicker evolution of the external shock of the afterglow to the time
when the jet becomes obvious, i.e. the jet-break time.  Thus this
observed variation among GRBs may result from a perspective effect due
to different observer angles of a morphologically homogeneous
populations of GRBs.

Also, a conjecture is made that peak luminosities not only vary
inversely with burst timescale, but also are directly proportional to
the spectral break energy.  If true, this could provide important
information for explaining the source of this break.

\end{abstract}

\date{\today}

\maketitle

\section{Introduction}

Only recently, with the discovery of afterglows and in turn,
redshifts for a handful of gamma-ray bursts, has there been progress
in trend spotting within the seemingly chaotic variety of gamma-ray
burst shapes and sizes.  \citet{nmb00} discovered an anti-correlation
between the isotropic peak gamma-ray luminosity, $L_{pk}$, of GRBs and
the pulse lag, $\Delta t$.  This lag is the time delay of the arrival
of a burst pulse in the BATSE detector low energy channels compared to
its arrival in the high energy channels.  Similarly \citet{frr00} and
also \citet{rlfr+00} have shown that a measure of the variability of
GRB lightcurves correlates with this peak luminosity.  Most recently
\citet{fksd01} have shown that the isotropic gamma-ray energy,
$E_{iso}$, is anti-correlated with the jet-break time, $\tau_j$.  The
jet-break time is when the afterglow lightcurve changes (typically
seen as a break) its decay rate, which is thought to be a
manifestation of the finite opening angle of the jet.

As demonstrated in \citet{sg01} these correlations are closely related
and are likely manifestations of the same physical effect.  As
discussed in the next section, we find an unexpectedly tight
relationship between spectral lags and jet-break times.  Thus we argue
that transitivity suggests that $L_{pk}$, $E_{iso}$, $\Delta t$ and
$\tau_j$ are all interrelated by power-laws.  In \citet{jay00,jay01}
it was argued that the lag-luminosity relationship, $L_{pk}$ vs.\
$\Delta t$, derives from kinematics: the variation in velocity of the
relativistic ejecta with respect to the observer.  In particular, the
Doppler factor, dependent upon the speed and angle of the emitter with
respect to the observer, will increase observed luminosity and
decrease observed timescales.  In \citet{sg01} we argue that all of
these relationships originate from kinematic variations among bursts.


\section{Discovery of a Correlation between Pulse Lags and Jet-Break Times}

\begin{figure}
\includegraphics[height=.35\textheight]{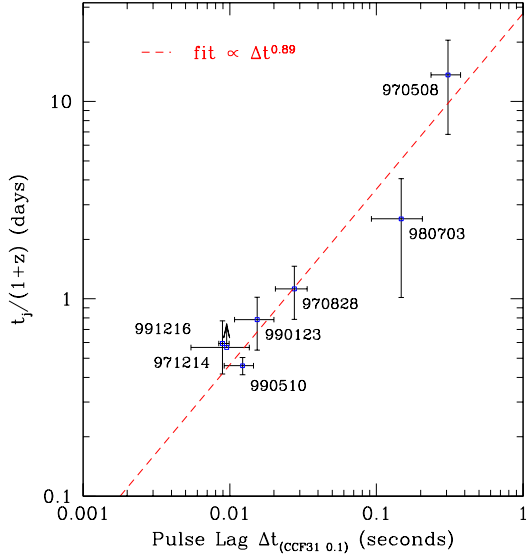}
\caption{Plot from \citep{sg01} of redshift-corrected burst pulse
lags, $\Delta t$, observed between BATSE channels 1 and 3, versus
observed jet-break times, corrected for redshift, $\tau_j \equiv
t_j/(1+z)$.  Jet break times $t_j$ are from \citet{fksd01} and pulse
lags are from \citet{nmb00}.  The fit, given by Eqn.~(\ref{t_jeqn}),
does not include GRB\,971214 which only has a lower limit on the
jet-break time.
\label{tjvtlag}}
\end{figure}

In \citet{sg01} we compare the two burst timescales: the redshift
corrected jet-break time, $\tau_j$, and the redshift corrected lags,
$\Delta t$.  We assembled a complete sample of seven bursts for which
there are data for $\Delta t$, $\tau_j$ and redshift $z$ (GRB\ 971214
has only a lower limit for $\tau_j$, so was not used in fits, but is
shown in the figures).  Using the CCF31 0.1 lags, $\Delta
t_{\text{(CCF31 0.1)}}$, determined by cross-correlating pulses in
BATSE channels 1 \& 3 down to 0.1 of the peak luminosity
\citep{nmb00}, a good fit results:
\begin{equation}
\tau_j = \frac{t_j}{1+z} = 28^{+ 18}_{- 11} 
{ \biggl(\frac{\Delta t_{\text{(CCF31 0.1)}} }{1~\text{sec}}\biggr)^{0.89 \pm 0.12} \text{days}}
\label{t_jeqn}
\end{equation} 
(shown in Fig. \ref{tjvtlag}) with a reduced chi-squared $\chi^2_r =
4.7/4$ and a respectable goodness-of-fit $Q = 0.31$ \citep{pftv88}.

The existence of such a close relationship between one timescale
associated with the GRB itself, and another timescale solely deriving
from the afterglow is surprising.  The standard GRB paradigm
\citep{piran00} says that the GRB derives from internal shocks in an
uneven relativistic wind, while the afterglow comes from a shock
sweeping into the ISM, obeying simple self-similar scaling laws and
thus not depending on initial conditions imposed by the GRB.  In
\citet{sg01} we discuss three possible models to explain the
relationship of Eqn.\ (\ref{t_jeqn}).

\section{Conjecture: Luminosity is Correlated to Break Energy} 

An enduring mystery in GRBs is the relative constance \cite{mppb+95}
of the observed break energy, $E_0$, of GRB spectra, represented by a
broken power-law ``Band function'' \citep{bmf+93}.  This mystery is
doubly troubling in light of the several relationships described
earlier in this paper.  How can $E_0$ be constant within a factor of
about three while timescales, luminosities, and energies vary over
almost two orders of magnitude?  Light might be shed on this issue
with the observation that there appears to be a correlation between
$L_{pk} \tau_j^\beta$ and $E_0$, where $\beta = 1.58$ is the index for
the $L_{pk}$ vs.\ $\tau_j$ power-law relationship (Eqn.\ \ref{Leqn})
found by \citet{sg01} .

\begin{figure}
\includegraphics[height=.35\textheight]{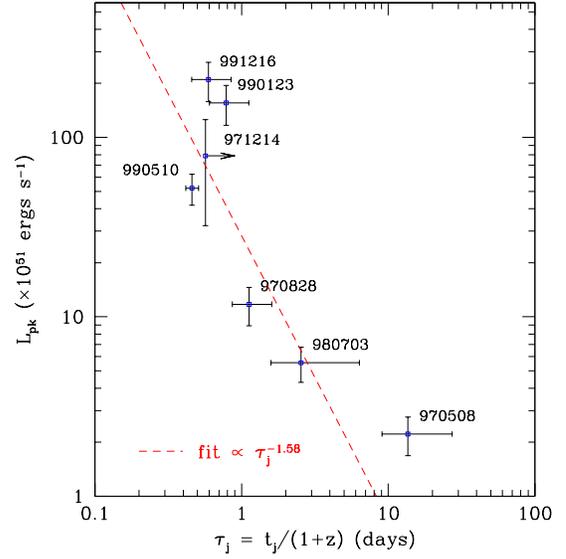}
\caption{Plot from \citep{sg01} of redshift-corrected burst peak
luminosites $L_{pk}$, versus redshift-corrected observed jet-break
times $\tau_j \equiv t_j/(1+z)$.  Jet break times $t_j$ are from
\citet{fksd01} and luminosities are calculated from \citet{jbp01}.
Because GRB\,971214 only has a lower limit on the jet-break time, it
is not included in the fit (given by Eqn.~\ref{Leqn}).
\label{Lvtlag}}
\end{figure}

This correlation is demonstrated by comparing the fit $L_{pk}$ vs.\
$\tau_j$ with that of $L_{pk}/E_0/(1+z)$ vs.\ $\tau_j$.  As in
\citet{sg01} we find

\begin{equation}
L_{pk} = 28^{+ 6}_{- 5} \times 10^{51} \Biggl(\frac{\tau_j}{1~ \text{days}}\Biggr)^{-1.58 \pm 0.23} \text{ergs s}^{-1} \label{Leqn}
\end{equation}  
(shown in Fig.~\ref{Lvtlag}) with $\chi^2_r = 29/4$ and $Q \sim
10^{-6}$.  While the correlation is plainly apparent, the fit is poor,
suggesting this relationship is not consistent with a simple
power-law.

\begin{figure}
\includegraphics[height=.35\textheight]{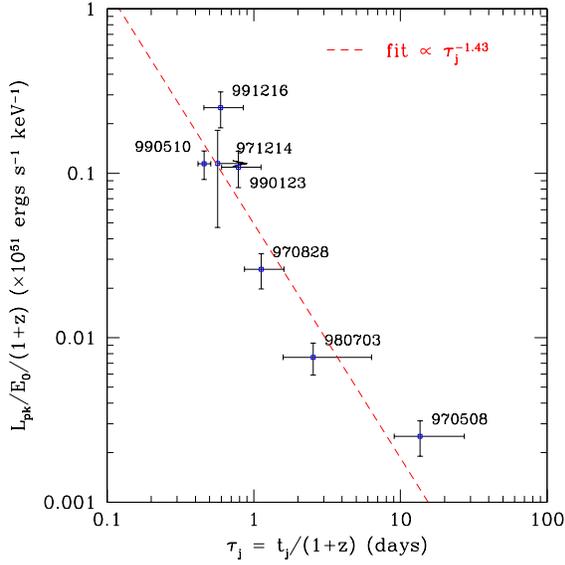}
\caption{Plot of redshift-corrected burst peak luminosites $L_{pk}$
divided by redshift corrected spectral break energies $(1+z) E_0$,
versus redshift-corrected observed jet-break times $\tau_j \equiv
t_j/(1+z)$.  The fit is given by Eqn.~(\ref{LoE0eqn}). Spectral break
energies, $E_0$, are from \citet{jbp01}.  See the caption of
Fig.~\ref{Lvtlag} for details.
\label{LoE0vtlag}}
\end{figure}

Now in order to compare with Eqn.~(\ref{Leqn}), I fit $L_{pk}/E_0/(1+z)$
vs.\ $\tau_j$ and find
\begin{equation}
\begin{split}
\frac{L_{pk}}{E_0 (1+z)} = &4.9\pm 8 \Biggl(\frac{\tau_j}{1~ \text{days}}\Biggr)^{-1.43 \pm 0.19}\\ & \times 10^{49} \text{ergs s}^{-1} \text{keV}^{-1} \label{LoE0eqn}
\end{split}
\end{equation}  
(shown in Fig.~\ref{LoE0vtlag}) with $\chi^2_r = 9.9/4$ and $Q =
0.04$.  The fit is substantially improved.  For the sake of
demonstration, the errors for $L_{pk}$ from Eqn.~(\ref{Leqn}) are used
in this fit.  Realistically one should also factor in the errors in
determining $E_0$.

This improvement in the fit of Eqn.\ (\ref{LoE0eqn}) over that of
Eqn.\ (\ref{Leqn}) leads one to hypothesize the existence of an
additional dependence on $E_0$ in the $L_{pk}$ vs.\ $\tau_j$
relationship.  Thus I propose
\begin{equation}
L_{pk} \propto E_0^\alpha \tau_j^{-\beta}
\label{LEteqn}
\end{equation}
where $\alpha$ and $\beta$ are both roughly unity.  This dependence
might be indicative of a mechanism for the spectral breaking
independent of the physical mechanism behind the relations described
previously.  One intriguing possibility is that it may indicate a
filtering of the gamma-rays that is responsible for the spectral
break.  As such, the more effective the filter, the lower the energy,
$E_0$, at which the spectrum is broken {\it and} the more attenuated
is the photon flux at all energies.  Such a filtering mechanism would
require both of these effects to account for the dependence suggested
in Eqn.~\ref{LEteqn}.  Future observations will be necessary to
confirm this conjecture, and future work will elaborate on its
physical cause.

This work was performed under the auspices of the U.S. Department of
Energy by University of California Lawrence Livermore National
Laboratory under contract W-7405-ENG-48.


\bibliographystyle{arlobib}
\def\apjl{ApJ}
\def\apj{ApJ}
\def\nat{Nature}
\def\physrep{Phys.~Rep.}

\begin{thebibliography}{12}
\newcommand{\enquote}[1]{``#1''}
\expandafter\ifx\csname natexlab\endcsname\relax\def\natexlab#1{#1}\fi
\expandafter\ifx\csname url\endcsname\relax
  \def\url#1{\texttt{#1}}\fi
\expandafter\ifx\csname urlprefix\endcsname\relax\def\urlprefix{URL }\fi

\bibitem[{{Norris} et~al.(2000){Norris}, {Marani}, and {Bonnell}}]{nmb00}
{Norris}, J.~P., {Marani}, G.~F., and {Bonnell}, J.~T. (\textbf{2000}).
  \enquote{Connection between energy-dependent lags and peak luminosity in
  gamma-ray bursts,} \apj, \textbf{534}, 248.

\bibitem[{{Fenimore} and {Ramirez-Ruiz}(2000)}]{frr00}
{Fenimore}, E.~E., and {Ramirez-Ruiz}, E. (\textbf{2000}). \enquote{Redshifts
  for 224 batse gamma-ray bursts determined by variability and the cosmological
  consequences,} astro-ph/0004176, submitted to ApJ.

\bibitem[{{Reichart} et~al.(2000){Reichart}, {Lamb}, {Fenimore},
  {Ramirez-Ruiz}, {Cline}, and {Hurley}}]{rlfr+00}
{Reichart}, D.~E., {Lamb}, D.~Q., {Fenimore}, E.~E., {Ramirez-Ruiz}, E.,
  {Cline}, T.~L., and {Hurley}, K. (\textbf{2000}). \enquote{A possible
  cepheid-like luminosity estimator for the long gamma-rayb bursts,}
  \apj, \textbf{552}, 57.

\bibitem[{{Frail} et~al.(2001){Frail}, {Kulkarni}, {Sari}, {Djorgovski},
  {Bloom}, {Galama}, {Reichart}, {Berger}, {Harrison}, {Price}, {Yost},
  {Diercks}, {Goodrich}, and {Chaffee}}]{fksd01}
{Frail}, D.~A., {Kulkarni}, S.~R., {Sari}, R., {Djorgovski}, S.~G., {Bloom},
  J.~S., {Galama}, T.~J., {Reichart}, D.~E., {Berger}, E., {Harrison}, F.~A.,
  {Price}, P.~A., {Yost}, S.~A., {Diercks}, A., {Goodrich}, R.~W., and
  {Chaffee}, F. (\textbf{2001}). \enquote{{Beaming in Gamma-Ray Bursts:
  Evidence for a Standard Energy Reservoir},} \apjl, \textbf{562}, L55--L58.

\bibitem[{{Salmonson} and {Galama}(2001)}]{sg01} {Salmonson}, J.~D., and {Galama}, T.~J. (\textbf{2001}). \enquote{Discovery of a tight correlation between pulse lag/luminosity and jet-break times: a connection between gamma-ray burst and afterglow properties.} \apj, \textbf{569}, 682-8. 

\bibitem[{{Salmonson}(2000)}]{jay00}
{Salmonson}, J.~D. (\textbf{2000}). \enquote{On the kinematic origin of the
  luminosity-pulse lag relationship in gamma-ray bursts,} \apjl, \textbf{544},
  L115--L117.

\bibitem[{{Salmonson}(2001)}]{jay01}
{Salmonson}, J.~D. (\textbf{2001}). \enquote{On the kinematics of grb 980425
  and its association with sn 1998bw,} \apjl, \textbf{546}, L29--L31.

\bibitem[{{Press} et~al.(1988){Press}, {Flannery}, {Teukolsky}, and
  {Vetterling}}]{pftv88}
{Press}, W.~H., {Flannery}, B.~P., {Teukolsky}, S.~A., and {Vetterling}, W.~T.
  (\textbf{1988}). \emph{Numerical Recipes in C: The Art of Scientific
  Computing} (Cambridge University Press).

\bibitem[{{Piran}(2000)}]{piran00}
{Piran}, T. (\textbf{2000}). \enquote{Gamma-ray bursts - a puzzle being
  resolved.} \physrep, \textbf{333}, 529--553.

\bibitem[{{Mallozzi} et~al.(1995){Mallozzi}, {Paciesas}, {Pendleton}, {Briggs},
  {Preece}, {Meegan}, and {Fishman}}]{mppb+95}
{Mallozzi}, R.~S., {Paciesas}, W.~S., {Pendleton}, G.~N., {Briggs}, M.~S.,
  {Preece}, R.~D., {Meegan}, C.~A., and {Fishman}, G.~J. (\textbf{1995}).
  \enquote{The nu f nu peak energy distributions of gamma-ray bursts observed
  by batse,} \apj, \textbf{454}, 597+.

\bibitem[{{Band} et~al.(1993){Band}, {Matteson}, {Ford}, {Schaefer}, {Palmer},
  {Teegarden}, {Cline}, {Briggs}, {Paciesas}, {Pendleton}, {Fishman},
  {Kouveliotou}, {Meegan}, {Wilson}, and {Lestrade}}]{bmf+93}
{Band}, D., {Matteson}, J., {Ford}, L., {Schaefer}, B., {Palmer}, D.,
  {Teegarden}, B., {Cline}, T., {Briggs}, M., {Paciesas}, W., {Pendleton}, G.,
  {Fishman}, G., {Kouveliotou}, C., {Meegan}, C., {Wilson}, R., and {Lestrade},
  P. (\textbf{1993}). \enquote{Batse observations of gamma-ray burst spectra. i
  - spectral diversity,} \apj, \textbf{413}, 281--292.

\bibitem[{{Jimenez} et~al.(2001){Jimenez}, {Band}, and {Piran}}]{jbp01}
{Jimenez}, R., {Band}, D., and {Piran}, T. (\textbf{2001}). \enquote{Energetics
  of gamma ray bursts,} \apj, \textbf{561}, 171.

\end{thebibliography}

\end{document}